\documentclass[manuscript]{aastex}

\usepackage{psfig,graphicx,multicol,color}
\usepackage{ulem, threeparttable}

\def\cm{{\rm\thinspace cm}}

\def\keV{{\rm\thinspace keV}}

\def\Lsun{\hbox{$\rm\thinspace L_{\odot}$}}

\def\Msun{\hbox{$\rm\thinspace M_{\odot}$}}

\def\ph{{\rm\thinspace ph}}
\def\s{{\rm\thinspace s}}

\def\powerlawfluxat1kev{\hbox{$\ph\cm^{-2}\s^{-1}\keV^{-1}$}}

\def\lapp{\ifmmode\stackrel{<}{_{\sim}}\else$\stackrel{<}{_{\sim}}$\fi}

\def\gapp{\ifmmode\stackrel{>}{_{\sim}}\else$\stackrel{>}{_{\sim}}$\fi}
\def\spose#1{\hbox to 0pt{#1\hss}}

\def\approxlt{\mathrel{\spose{\lower 3pt\hbox{$\sim$}}
        \raise 2.0pt\hbox{$<$}}}
\def\approxgt{\mathrel{\spose{\lower 3pt\hbox{$\sim$}}
        \raise 2.0pt\hbox{$>$}}}

\def\lapp{\ifmmode\stackrel{<}{_{\sim}}\else$\stackrel{<}{_{\sim}}$\fi}
\def\gapp{\ifmmode\stackrel{>}{_{\sim}}\else$\stackrel{>}{_{\sim}}$\fi}

\def\mcg6{MCG$-$6-30-15}
\def\1624{4U~1624$-$490}
\def\mrk766{Markarian~766}
\def\mr2251{MRC~2251-178}
\def\ngc2110{NGC~2110}
\def\ir13349{IRAS~13349+2438}
\def\iras18325{IRAS~18325--5926}
\def\grs1915{GRS~1915+105}
\def\cirx1{Cir~X-1}
\def\xtej1748{XTE~J1748-288}
\def\chandra{{\it Chandra }}

\def\suzaku{{\it Suzaku }}
\def\XMM{{\it XMM-Newton }}
\def\xmm{{\it XMM }}

\def\xtegammamcg6{$\Gamma=1.9$}

\def\ovii{O~{\sc vii }\,}

\def\fe25{Fe~{\sc xxv}\,}
\def\fe26{Fe~{\sc xxvi}\,}
\def\Ne9{Ne~{\sc ix }\,}
\def\ne10{Ne~{\sc x }\,}
\def\mg11{Mg~{\sc xi }\,}
\def\si13{Si~{\sc xiii }\,}

\def\apj{ApJ}

\def\apj{ApJ}
\def\apjs{ApJS}
\def\aap{A\&A}

\def\ovii{{\sc O~vii}$\,$}

\def\si4{Si~{\sc iv}}
\def\fe25{Fe~{\sc xxv}}
\def\fe26{Fe~{\sc xxvi}}
\def\mg2{Mg~{\sc ii}}
\def\Msun{\ifmmode M_{\odot} \else $M_{\odot}$\fi}
\def\Lsun{\ifmmode L_{\odot} \else $L_{\odot}$\fi}

\tighten

\begin{document}

\shortauthors{Lee et al.}
\shorttitle{An X-ray technique for determining the composition and quantity of interstellar dust}


\title{Condensed Matter Astrophysics \\ A Prescription for Determining the Species-Specific Composition and Quantity of Interstellar Dust  using X-rays}

\author{Julia C. Lee, Jingen Xiang}
\affil{Harvard University, Department of Astronomy\footnote{a part of the Harvard-Smithsonian Center for Astrophysics} \\
Harvard-Smithsonian Center for Astrophysics, 60 Garden Street MS-6, Cambridge, MA 02138}
\and
\author{Bruce Ravel}
\affil{National Institute of Standards and Technology \\
100 Bureau Drive, Gaithersburg, MD 20899}
\and
\author{Jeffrey Kortright}
\affil{Lawrence Berkeley National Laboratory, Materials Sciences Division \\ 1 Cyclotron Road MS-2R0100, Berkeley, CA 94720 }
\and
\author{Kathryn Flanagan}
\affil{Space Telescope Science Institute\\ 3700 San Martin Dr Baltimore MD 21218
}





\begin{abstract} We present a new technique for determining the
{\it quantity and composition} of dust in astrophysical environments using
$<$~6~keV X-rays.  We argue that high resolution X-ray spectra as
enabled by the Chandra and XMM-Newton gratings should be considered a
powerful and viable  new resource for delving into a relatively
unexplored regime for directly determining dust properties:
composition, quantity, and distribution.  We present initial
cross-section measurements of astrophysically likely iron-based dust
candidates  taken at the Lawrence Berkeley National Laboratory
Advanced Light Source synchrotron beamline, {as an illustrative
tool for the formulation of our technique for determining the  quantify and composition
of interstellar dust with X-rays.   (Cross sections for the
materials presented here will be made available for astrophysical modelling
in the near future.)}
Focused at the 700~eV
Fe~$\rm L_{\sc III}$ and $\rm L_{\sc II}$ photoelectric edges, we
discuss a technique for modeling dust properties  in the soft X-rays
using L-edge data, to complement K-edge X-ray absorption fine
structure analysis techniques discussed in \cite{leeravel:xafs05}.
This is intended to be {\it a techniques paper} 
of interest and usefulness to both
condensed matter experimentalists and astrophysicists.   For the
experimentalists, we offer a new prescription for normalizing
relatively low S/N L-edge cross section measurements.   For
astrophysics interests, we discuss the use of X-ray absorption spectra
for determining dust composition in cold and ionized astrophysical
environments, and {\it a new method for determining
species-specific gas-to-dust ratios}. Possible astrophysical applications of
interest, including relevance to Sagittarius A$^\ast$  are offered.
Prospects for improving on this work in future
X-ray missions with higher throughput and  spectral resolution are
also presented in the context of spectral resolution goals for 
gratings and calorimeters, for proposed and planned missions such as
Astro-H and the International X-ray Observatory.

\end{abstract}

\section{Introduction}
\label{sec:intro}  Understanding dust is vital to our understanding of
the Universe. Dust is a primary repository of the interstellar medium
(ISM) and contributes to the chemical  evolution of stars,
planets, and life itself. A better understanding of the content of
astrophysical dust has  far reaching relevance and application to
areas of astrophysics ranging from nucleosynthesis to planet
formation.    Dust is everywhere, and for many astrophysical topics,
its presence hinders our studies in some way, be it to extinguish a UV
part of a spectrum, or possibly affect Cosmology results, as e.g., in
the case of using Type~Ia supernovae light curves (which are affected
by line-of-sight dust) as a beacon for probing the dark energy content
in the Universe.   Therefore, at a minimum, a better understanding of
dust  will allow us to better isolate its effects (be it complicating
spectra or light curves) to get at a cleaner study of topics from
cosmology to black hole environments.  There is a multi-wavelength
industry (mostly radio to IR) focused on the problem of astrophysical
dust, and in recent years, Spitzer observations have significantly
improved our understanding.  Yet, despite good
progress, there remains much to understand of dust properties (size,
composition, and distribution) in astrophysical environments,  from
the colder ISM environs to the hotter environments near the disks and
envelopes of young stars and compact sources.  Our quest to address
outstanding problems, therefore, can greatly  benefit from additional
complementary and/or orthogonal techniques. 

There are several advantages to studying dust properties in the
X-rays, the most significant of which is that we can {\it directly}
measure its quantity (relative to gas phase), {\it and} composition.
(``Dust" in this papers includes complex molecules).  In the UV and
optical,  the presence of dust is generally inferred through
depletion, rather than directly measured.   Radio spectroscopy is able
to probe the presence of certain molecules, and IR, some subset of ISM
compounds (mostly polycyclic aromatic hydrocarbon or PAHs, graphites,
certain silicates, and ice mantle bands).  In these bands, spectral
features come from probing the molecule as a whole through  rotational
or vibrational modes originating from e.g.  the  excitation of phonons
(rather than electrons).   Both gas and ($\approxlt 10\mu$m) dust
are semi-transparent to X-rays, so that the measured absorption in this
energy band is sensitive to \textit{all} atoms in both gas and solid
phase.  Therefore, these high energy photons can be used to facilitate
a direct measurement of condensed phase chemistry via a study of element-specific
 atomic processes, whereby the excitation of an electron to a higher lying
quantum level, band resonance, or molecular structure will imprint
modulations on X-ray spectra that reflect the individual atoms which
make up the molecule or solid.  Therefore, in very much the same way we
can look at an absorption or emission line at a given energy to
identify ions,  the observed spectral modulations near photoelectric
edges, known as X-ray absorption fine structure (XAFS) provide unique
signatures of the condensed matter that imprinted that signature.  For
this reason, high resolution X-ray studies, as currently best enabled
by X-ray grating observations (Chandra and XMM), provide a unique and
powerful tool for  determining the state and composition of ISM
grains.

\cite{leeravel:xafs05} have already discussed the viability of XAFS
analysis and techniques for K-edge data with particular emphasis on
iron in the hard $> 7$~keV X-rays, in early anticipation of the launch
of the \suzaku calorimeters.  (See also \citealt{astroxafs95},
\citealt{astroxafs97}, and \citealt{astroxafs98} for early theory
discussions relating to XAFS detections in astrophysical environments.)
Due to the unfortunate in-flight failure of the instrument, the
viability of astrophysical XAFS studies  will have to focus on the
softer $<$~6~keV regime with existing instruments, until the Astro-H
launch.   While a more complicated part of the spectrum
(due to the numerous ions which populate this spectral region), XAFS
signatures in the soft X-rays have been reported  in early papers based on \chandra
and \xmm spectral studies  (\citealt{xrb_paerels:01,jcl_mcg6_wa1:01,
jcl_grs1915:02, xafs_ueda:05,crabdust:09,scox1xafs09}; see also
\citealt{ism_takei:02,schulz_cygx1:02,juett_ism:04,juett_ism:06} for studies
focused on the oxygen~K edge for abundance determinations.)  It is the
detection of these  features in X-ray spectra of astrophysical objects
which have motivated us to obtain laboratory measurements of likely
ISM grain candidates to facilitate astrophysical modeling efforts.  To
date, no modelling of the magnitude we propose here has been
undertaken.  The $\sim 0.1-8$~keV energy coverage combined between the
\chandra and \XMM gratings will allow us to look in more detail at
photoelectric edges near C~K, O~K, Fe~L, Mg~K, Si~K, Al~K,  S~K, Ca~K,
and Fe~K, and therefore molecules/grains containing these constituents.  We
note however that for the K-edges of C, S, Ca and Fe, we are likely to
be limited by signal-to-noise (hereafter, S/N) and/or spectral
resolution,  but are nevertheless well situated to study magnesium-,
silicate, and aluminum-based grains using K-edge data, and iron-based
grains/molecules can be studied using L-edge spectra at $\sim 700$~eV.
Additionally, because X-ray  absorption spectra is from some admixture of gas and dust,
each having very different individual spectral signatures, we can
separate their relative contributions through spectral modelling.
Our proposed technique for doing so is discussed in
\S\ref{sec:gasfromdust}, {using  cross-section
 measurements of XAFS at Fe~L as the illustrative example for our discussion points.}

The intent of this paper is to present dialog focused on the
application of  condensed matter {\it X-ray} techniques  to the soft
X-ray {\it absorption} study of astrophysical dust properties:
composition and quantity.  What  knowledge we gain from this work will
greatly complement current IR studies of dust.  Ultimately, the
questions we would strive to address over the course of this study
include: (1)~the mineralogy of dust in different astrophysical
environments, and perhaps (2) something of its nature
(e.g. crystalline or amorphous).  Since K-edge analysis has already
been discussed by \cite{leeravel:xafs05}, we focus our attention here
on XAFS studies as applied to L-edges, where the spectral modeling of
{\it near}-edge absorption modulations (XANES: X-ray absorption near
edge structure), rather than {\it far}-edge  absorption modulations
(EXAFS: Extended X-ray absorption fine structure) play a primary role.
While the emphasis is on the $\sim 700$~eV Fe~L spectral region, the
techniques presented will be widely applicable to all L-edge studies.
Because condensed matter measurements are discussed in the context of
astrophysics studies, we organize this paper into sections that would
be of interest and usefulness to both condensed matter experimentalists
(\S\ref{sec:lab}), and astrophysicists (\S\ref{sec:gasfromdust},
\S\ref{sec:discussion}), For the experimentalists, \S\ref{sec:norm}
may be of particular interest for its discussion of an L-edge cross
section normalization technique that is robust to laboratory
synchrotron cross-section measurements with relatively poor S/N.  For
astrophysics interests, \S\ref{sec:gasfromdust} discusses a new method
for determining interstellar dust quantity and composition, through
X-ray studies.  This is followed in \S\ref{sec:discussion} with a
brief discussion of  some possible studies of interest, and thought
experiment discussion on this technique's ability to separate out
fractional contributions of different composition dust in different
regions.  Prospects for improving on this work in the context of
future missions is discussed in \S\ref{sec:future}.

\section{The samples and laboratory experiment} 
\label{sec:lab}   UV, IR, and planetary studies of meteorites and dust
have pointed to interstellar dust grains condensed from heavier
elements such as C, N, O, Mg, Si, and Fe.  With respect to using
\chandra High Energy Transmission Grating Spectrometer (HETGS) spectra to enable detailed studies of absorption
structure near photoelectric edges to determine grain composition,
Fe-, Mg-, and Si-based dust would be the most relevant since the
L-edge of Fe and K-edges of Mg and Si are all contained within the
$\sim 0.5-8$~keV ($\sim 1.5-25$\AA) HETGS bandpass, where the
effective area is also highest. (The Fe K-edge is also encompassed by
the HETGS, but the S/N there is usually too low to be useful for
analysis.)  For this paper, we compare cross section measurements for
hematite ($\rm \alpha-Fe_2O_3$), iron sulfate ($\rm FeSO_4$),
fayalite ($\rm Fe_2SiO_4$), and lepidocrocite ($\rm \gamma-FeOOH$).
{ Cross-section measurements at Fe~L and Si~K 
are nearing completion
for the most astrophysically-likely iron- and silicon-based dust,
which will be presented in a future paper on XAFS standards for
astrophysical use, while the data presented in this paper is used 
primarily to illustrate our proposed analysis techniques and methodology. }
Planned measurements
are also being made for other interesting edges discussed previously.

The laboratory data presented in this paper were taken at the
Lawrence Berkeley National Laboratory Advanced Light Source
beamline~6.3.1.  This is a bending magnet beamline with a Variable
Line Spacing Plane Grating Monochromator (VLS-PGM) that is sensitive
to 300-2000~eV energy range and has a resolving power $\rm E/\Delta E
= 3000$. This  resolution exceeds the $\rm E/\Delta E =1000$ \chandra
HETGS resolving power by factor of three. 

The 6.3.1 detectors allow for choice between total electron yield,
photodiode (fluorescence), and transmission experiments.  The latter
would have been the ideal choice, since transmission measurements
would be directly related to what we seek to understand, namely the
transmission of X-rays through interstellar grains of comparable
thickness.  However, a nominal thickness estimate for 30\%
transmission would require film/foil thicknesses for our candidate
samples in the range $\rm t \sim 0.1-0.8 \mu m$ from the transparent
to opaque side of the edge at Fe~L, as calculated according to the
formalism $T = e^{-\mu \rho t}$.  (The density $\rho$ for the
compounds were taken from ``The Handbook of Chemistry and Physics'',
and the values for the attenuation length were obtained from the
\footnote{http://www-cxro.lbl.gov/optical\_constants/}CXRO at LBL, and
\footnote{http://physics.nist.gov/PhysRefData/FFast/html/form.html}NIST.)
 Preparation of pure-phase samples of that thickness was 
impractical. As such, we opted to obtain cross-section 
measures using both fluorescence and electron yield (\S\ref{sec:experiment}). 

\subsection{Fluorescence and Electron Yield Experiments} 
\label{sec:experiment}  In an X-ray-absorption spectroscopy (XAS) experiment, the incident X-ray
photon promotes a deep-core electron into an unoccupied state above
the Fermi energy.  For the iron L edge experiments shown in this
paper, that deep core state is a $2p$ state.  As the incident photon
energy is varied using a variable line spacing plane grating
monochromator, the cross section of this absorption process is
monitored by measurement of the decay of a higher-lying electron into
the core-hole.  The fluorescence of a photon and emission of Auger
and secondary electrons can be monitored in tandem.  Because the mean free path of
the electron is limited to a few nanometers or less, the
electron yield measurement is mostly sensitive to the surface of the
sample. The fluorescent photon, on the other hand, has a larger mean-free path,
and therefore is a better probe of the bulk of the sample.

The samples measured in this paper were fine powders, either purchased
as such or prepared by grinding in an agate mortar and pestle.  A
portion of these fine powders were sprinkled onto conducting carbon
tape, another portion was pressed onto a flattened strip of indium
metal.  The carbon tape and indium strips were affixed to a glass
microscope slide for mounting onto the sample holder.  The fluorescence
measurement is made using a photodiode pointed at the sample.
However, since all of our samples were thick compared to the
penetration depth and highly concentrated in iron, the fluorescence
data were significantly attenuated by the self-absorption effect (see
e.g. \citealt{boothbridges05}).  All data in this paper were measured
in electron yield mode.

To measure the electron yield, an alligator clip was clamped onto the
glass slide and electrical contact was made with the otherwise isolated
carbon tape and indium strip, and used to carry the photocurrent into a current-to-voltage amplifier.  This
experimental arrangement provided for redundant measurements of each
sample.  Some samples proved more amenable to measurement on the
carbon tape substrate and others to the indium.  For every sample,
each measurement was repeated three times.  In each case, the
measurement yielding the highest quality, most reproducible data 
was selected for presentation.  The three scans for each sample have
been calibrated in energy, aligned, and averaged.

Absolute energy calibration was attained by measuring an iron foil and
calibrating it to the tabulated energies of the iron $L_{\rm II}$ and 
$L_{\rm III}$ edges \citep{CL-abstables} .  We had no capacity to prepare the surface of our
foil, thus we relied upon the fluorescence measurement.  Although the
foil measurement is severely attenuated by self-absorption, the edge
position can be reliably extracted.  Iron edge scans on other
materials were aligned to the iron foil data using reproducible
features in the spectrum of the incident intensity.

\subsection{Absorption cross section normalization technique} 
\label{sec:norm}   
In an XAS measurement, the absolute scale of the measured cross
 section is ambiguous.  The size of the measured step at an
 absorption edge energy depends on a wide variety of factors,
 including the concentration of the absorbing atom in the sample, the
 electronic gains on the signal chains, and the efficiencies of the
 detectors.  In order to compare measurements on disparate samples
 and under differing experimental conditions, it is standard practice
 to perform a common normalization of every measured spectrum.  In
 this work, we choose to perform this normalization by matching the
 measurements to tabulated values for bare-atom cross sections, as
 described in this section.  In this way, we can directly and
 consistently compare measurements on our samples to one another as
 well as to astronomical observations.

\begin{figure}
\includegraphics[width=0.75\textwidth,angle=270]{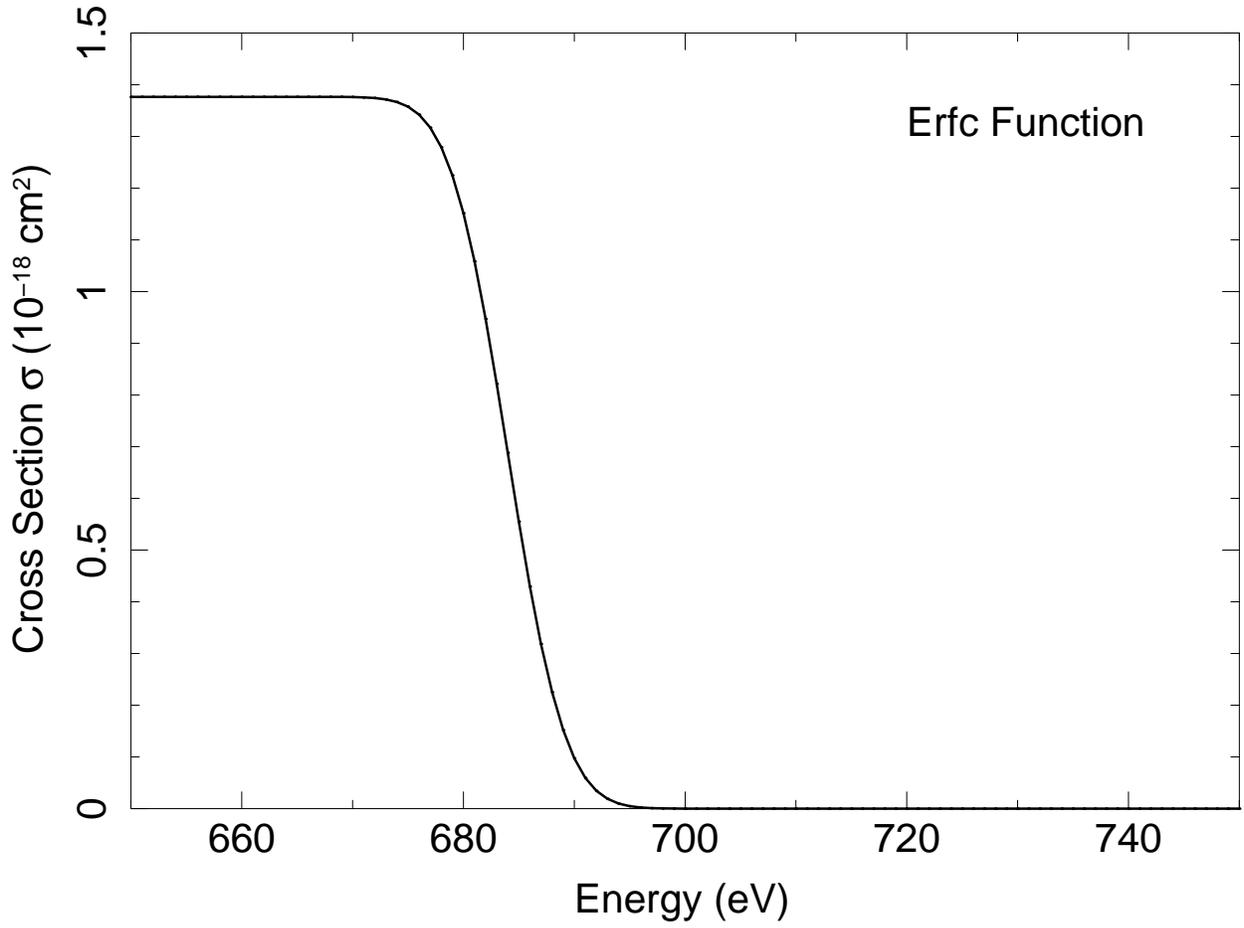}
\caption{The error function as defined in Eq.~\ref{eq:bkg},
which accounts for any decreasing function in the pre-edge slope,
due to residual inelastic scattering.  
\label{fig:erfc}}
\end{figure}

\begin{figure*}
\includegraphics[width=0.60\textwidth,angle=0]{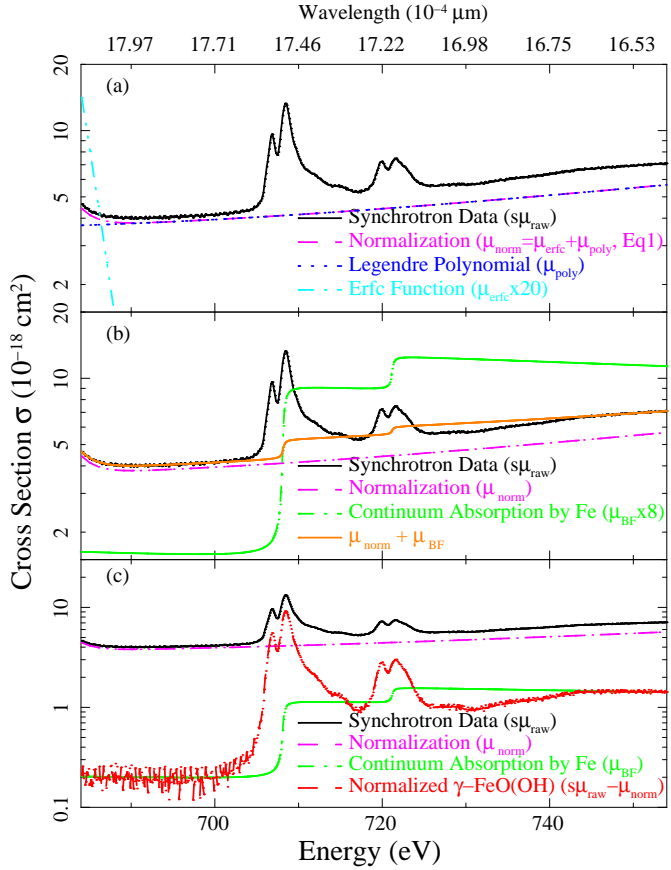}
\caption{An illustration, based on $\gamma$-FeO(OH) of the steps involved in the normalization of 
Fe~L edge synchrotron data as detailed in \S\ref{sec:norm}, and Eqs.~\ref{eq:bkg}~and~\ref{eq:norm}.
   (a) Our measured cross section scaled by $s$ (black), plotted with
the best determined normalization $\mu_{norm}$ (magenta), as defined in 
 Eq.~\ref{eq:bkg} to be the sum of $\mu_{\rm poly}$ (dashed dark blue) and
$\mu_{\rm erfc}$  (dashed-dot light blue).
(b)  The pre- and post-edge regions are then normalized against the 
tabulated cross sections $\mu_{\rm CL}$ of \cite{CL-abstables} which are shown
here as a step function arbitrarily enhanced to $ 5\times$ its value (dash-dot green), 
for illustrative purposes.  
Shown in orange solid is the $\mu_{\rm CL}$+$\mu_{\rm back}$ component
of the Eq.~\ref{eq:norm} minimizing function.
(c)  The {\it  normalized} $\gamma$-FeO(OH) cross sections (red) representing
bound-bound absorption,  plotted with the $\mu_{\rm CL}$  absorption cross-sections 
(green-dashed) respresenting bound-free  continuum absorption by Fe~L;
also plotted is the pre-normalization data of panel-a (black).
\label{fig:bkg}}
\end{figure*}

The normalization of XAS L edge data is more complicated than for K edge
data due to the close proximity of the $L_{II}$ and $L_{III}$ edges  in the soft X-ray
regime.  In order to properly normalize these spectra, we adopt the
{\it MBACK} normalization technique of \cite{weng05:softXnorm}, 
{\it with modifications}.  Here, we describe
the {\it MBACK} technique in brief, and our modification, which we
found to be necessary for data with $\approxgt$~0.2\% noise, as
typical of each individual cross section measure. We note that the
typical practice has been to first  co-add all individual
cross-section measures to increase S/N, before normalization.
However, we wished to be more rigorous in our methodology by first
renormalizing the individually measured  cross-sections before
combining to create the final used for fitting.  The {\it MBACK}
technique ensures better normalization  for the more complex
L-edge cross sections by calculating a smooth normalizing function over
the entire measured data range, rather than extrapolating from
independently pre-  and post-edge linear fits as would be sufficient
for the less complex K-edge cross sections.  For our data, this
normalizing function, $\mu_{norm}$, is dominated by the Legendre polynomial term
(in brackets) of the function:
\begin{equation} 
\mu_{norm}(E_i) = \left[\sum_{j=0}^{m}C_{j}(E_i-E_{edge})^j \right] + A\,\cdot {\rm erfc}({{E_i-E_{em}}\over\xi})
\label{eq:bkg}
\end{equation} 
where $E_{edge}$ is the energy of the onset of the edge-absorption,
and $m$ and $C_j$ are respectively the polynomial order and
coefficient.  Weng et al. additionally note that a rapidly decreasing
pre-edge shape, due to residual  elastic scattering is often seen in
fluorescence experiments using energy discriminating detectors, 
and therefore include an additional error
function component in the normalization  as defined by the non-bracketed
term of Eq.\ref{eq:bkg}$-$ see Fig.~\ref{fig:erfc} for the shape of
this function as determined from our sample fit to
$\gamma$-FeO(OH)$-$lepidocrocite.  Here, $E_{em}$ and $\xi$ are  respectively the
centroid energy of the X-ray emission line of the absorbing element,
and its width, with $A$ as a scale factor.  
We note however that this function is adopted here {\it solely}
as an effective way to model similar pre-edge feature shapes
originating from experimental artifacts of uncertain origin.
(Residual  elastic scattering effects do not apply to our experiment,
which measured total photocurrent.)
In any case, our data do not show a
large rapidly decreasing pre-edge slope so this term has negligible
effect on our overall normalization; nevertheless, we include it in
our modelling efforts for  completeness, since this would allow for a
global prescription for the determination of $\mu_{norm}$ that  is
independent of experimental technique. -- See Fig.~\ref{fig:bkg}a for
an illustration of the components which make up the normalizing function for our
sample $\gamma-$FeO(OH) compound.  In general, the details of the  normalization
variables would clearly depend on the  composition of the condensed
matter.

\begin{figure}
\includegraphics[width=0.80\textwidth,angle=0]{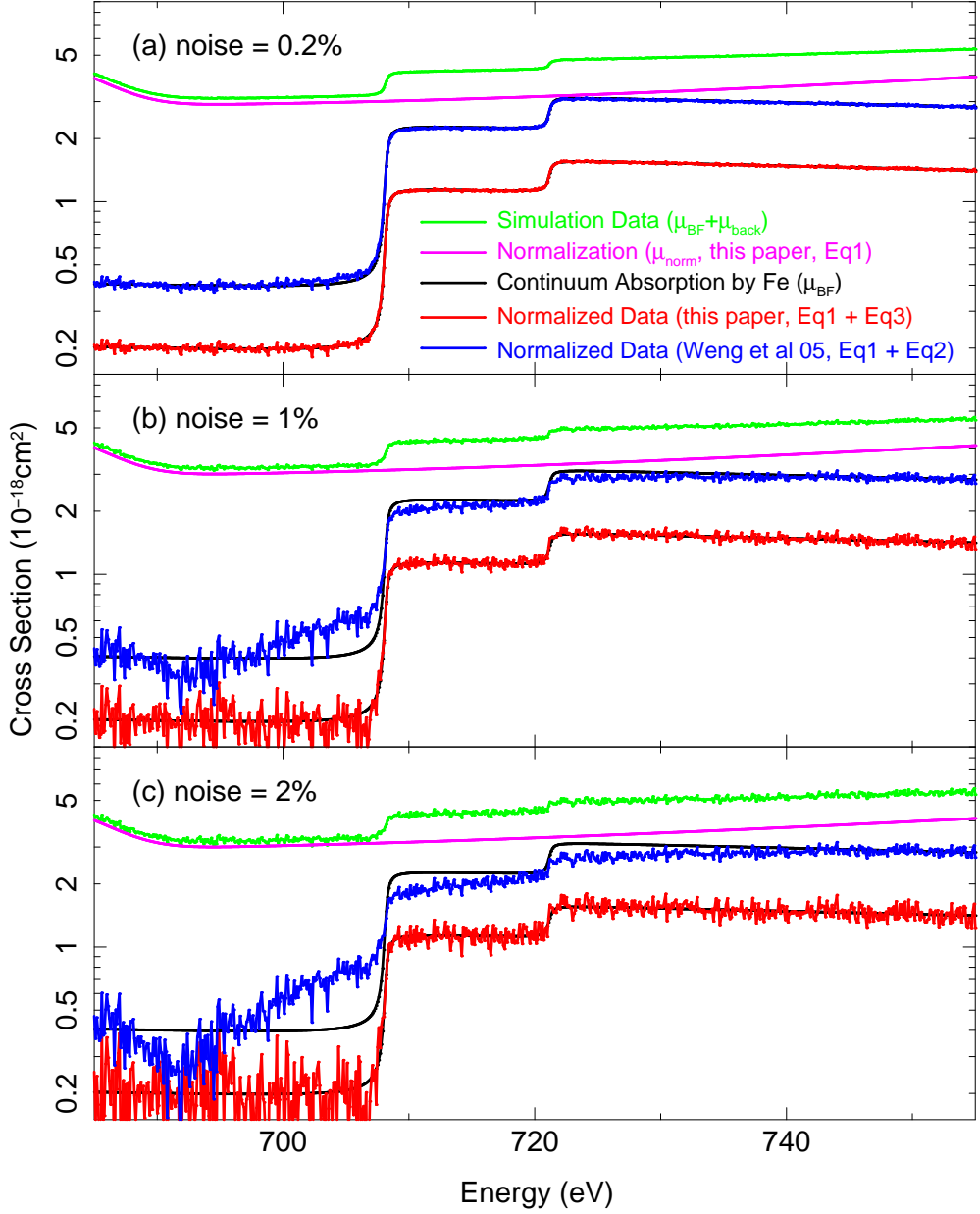}
\caption{A comparison of the normalization technique for
compound $\gamma-$FeOOH, 
based on Eq.~\ref{eq:wengnorm} (normalized data
based on the method of \citealt{weng05:softXnorm} shown
in blue; arbitrarily scaled up for illustrative purposes), 
versus its modification (Eq.~\ref{eq:norm}) as proposed in this paper 
(normalized data shown in red)
for simulated data (green) with varying levels of 
Gaussian distributed noise.   The Fe~L cross sections of 
\cite{CL-abstables} are shown in black to illustrate goodness of fit.
The normalization function of Eq.~\ref{eq:bkg}  is show in magenta. 
\label{fig:norm_sim}}
\end{figure}

To determine a best normalization (Fig.~\ref{fig:bkg}b), we {\it
initially} employ the prescribed minimizing function of Weng et al.:
\begin{equation}
\label{eq:wengnorm} 
{1\over n_1} \sum_{i=1}^{n_1}[\mu_{bf}(E_i)+\mu_{norm}(E_i)-s\mu_{raw}(E_i)]^2\\ +{1\over{n_2}}
\sum_{i=N-n_2+1}^{N}[\mu_{bf}(E_i)+\mu_{norm}(E_i)-s\mu_{raw}(E_i)]^2
\end{equation} 
where $N$ represents the total number of data bins, $i$
is the bin index, and $n_1$ and $n_2$ are respectively the total
number of pre-edge and post-edge data bins, such that e.g. $N-n_2+1$
represents index~1 of the post-edge bin. The  variable $\mu_{bf}$ is
the \cite{CL-abstables} tabulated absorption cross-sections representing
the bound-free non-resonant absorption component of Fe.
$\mu_{raw}$ is our measured cross-sections which is scaled by $s$, and $\mu_{norm}$ is the
normalization function of Eq.~\ref{eq:bkg}.  However,  in our fitting
efforts, we found the scale factor $s$ to be very  sensitive to noise,
often taking on small values for our data which  resulted in badly
normalized cross sections.  In considering a remedy for this, we
tested Eq.~\ref{eq:wengnorm} based on simulated cross sections
($\mu_{bf} + \mu_{norm}$) where we have added controlled levels of
Gaussian distributed noise, as per the prescription 
$(1 + f\, G_{noise})\cdot(\mu_{bf} + \mu_{norm})$, where $G_{noise}$ is 
a group of random numbers following the  Gaussian distribution,
and modified  by the fractional level $f$, i.e. $f=0.2\%$ or $f=1\%$,
as in Fig.~\ref{fig:norm_sim}.  (The Gaussian distribution function
is defined as $\phi(x)={e^{-(x-\mu)^2/2\sigma^2}} / \sigma \sqrt{2\pi}$
where the expectation value  $\mu=0$ and the  standard deviation $\sigma=1$.)
(The $\mu_{norm}$ parameters of
Eq.~\ref{eq:bkg} are set to $C_0$ = 3.0, $C_1$ = 0.01, $C_2$ = 0.0002,
$A$ = 1.0, $E_{em}$ = 685~eV, $\xi$ = 5.0, $E_{edge}$ = 707~eV).  We
then divide the simulated data by the value of $s$ arbitrarily set to
30, before fitting according to the minimization function of
Eq.~\ref{eq:wengnorm} to see if we can recover the same value.  In
doing so, we find that the Weng et al. method, {\it sans
modification}, is only robust for data with noise at
$\approxlt$~0.2\%. Accordingly, we modify Eq.~\ref{eq:wengnorm} by
dividing $s\mu_{raw}(E_{i})$ into the pre-edge and post-edge
minimization terms as per:
\begin{equation}
\label{eq-new} 
{1\over n_1} \sum_{i=1}^{n_1}\left[{\mu_{tab}(E_i)+\mu_{norm}(E_i)-s\mu_{raw}(E_i)
\over s\mu_{raw}(E_i)} \right]^2 +{1\over n_2}
\sum_{i=N-n_2+1}^{N}\left[{\mu_{tab}(E_i)+\mu_{norm}(E_i)-s\mu_{raw}(E_i)
\over s\mu_{raw}( E_i)}\right]^2
\label{eq:norm}
\end{equation} 
and find that our method (Eq.~\ref{eq:norm}) is robust
to data of up to 2\% noise.  In total, the normalization procedure
will use $m+5$ adjustable parameters ($m+1$ polynomial coefficients,
two scale factors of $s$ and $A$ as well $\xi$ and $E_{em}$) to
determine the best normalization for the measured cross sections.

\begin{figure}
\includegraphics[height=0.70\textheight, angle=0]{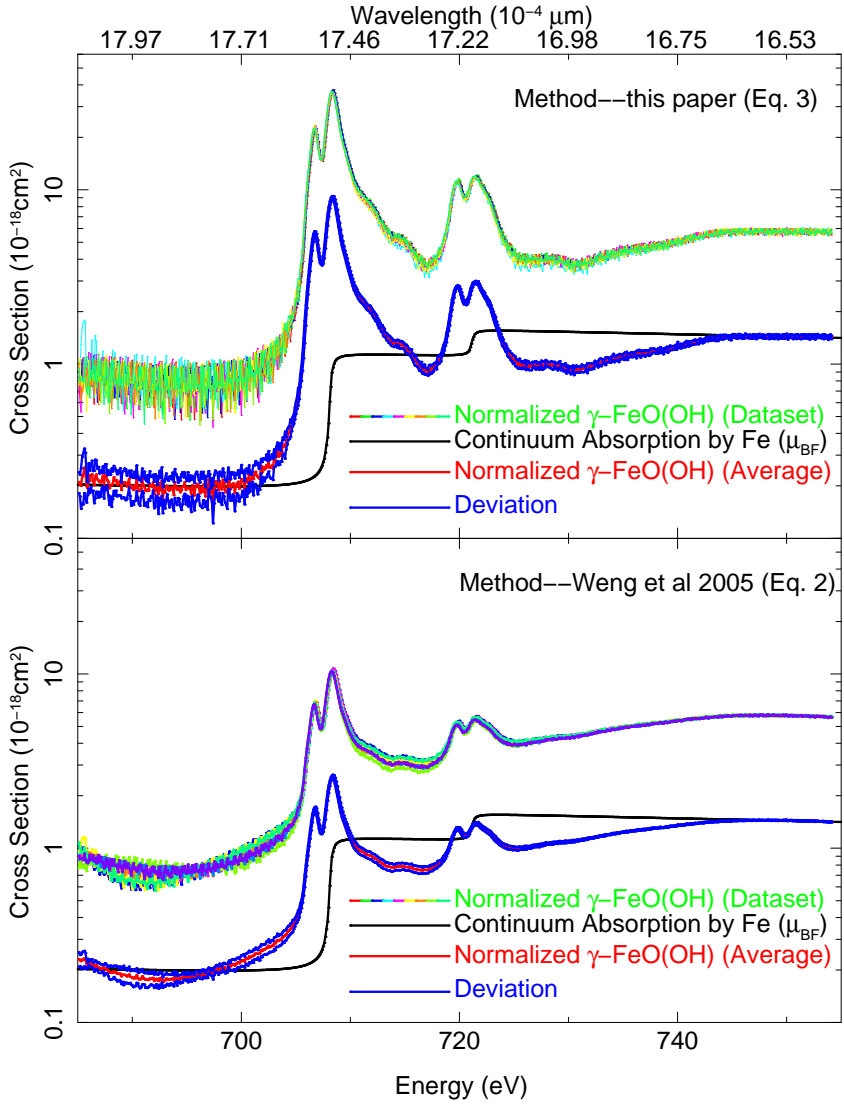}
\caption{A comparison of the cross section normalization technique 
proposed in this paper (top; Eq.~\ref{eq:norm}) versus that proposed 
by \citealt{weng05:softXnorm} (Eq.~\ref{eq:wengnorm}; bottom)
for the sample compound $\gamma-$FeO(OH).  In {\it each} panel, individual 
cross section measurements (arbitrarily scaled higher to facilitate
 plotting purposes) are shown in multicolors, while directly below,
the combined normalized data according to the respective methods
are shown in red. The blue bracketing this is the standard deviation 
based on the individual measurements to give sense of measurement
differences as a function of energy.  The \cite{CL-abstables}
cross sections are shown in black to compare how the normalization techniques
differ. 
\label{fig:normcompare}}
\end{figure}

\begin{figure*}
\includegraphics[height=0.55\textheight, angle=0]{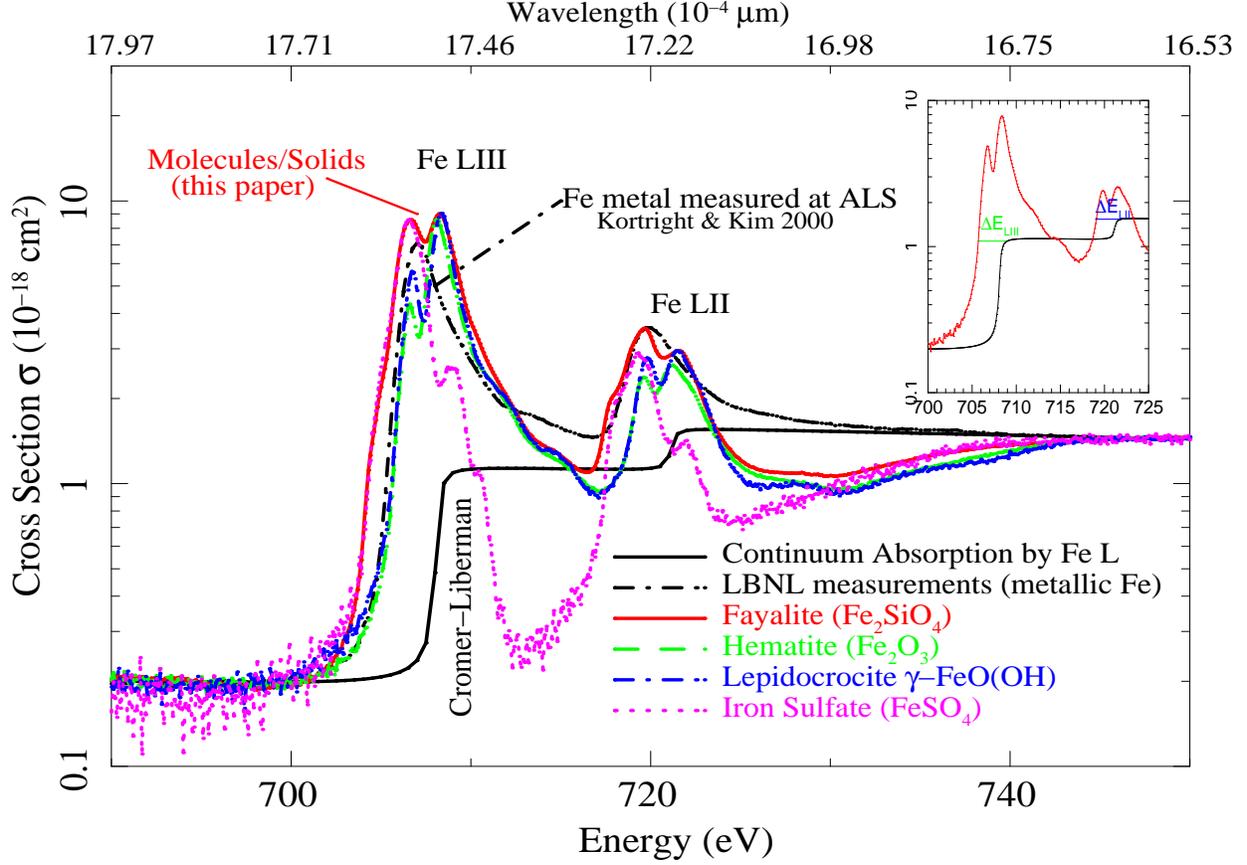}
\caption{X-ray absorption near edge structure in the vicinity of the
Fe~L photoelectric edge, post normalization, reveal that 
structure known as XAFS are distinct for different states
of condensed matter.   Notice also differences in edge structure between
bound-free continuum absorption (solid black step function) versus metallic (dashed
black) and molecular (in color) states.  Note that these are only
preliminary measurements which we intend to improve upon before
incorporation into astrophysical databases for common use.
The inset illustrates the $\Delta E$ values distinguishing
condensed matter (red) from gas-phase (black) absorption
at the Fe~$\rm L_{III}$ and $\rm L_{II}$ photoelectric edges energies,
for the different compounds which are tabulated in Table~\ref{tab:feledges}.
\label{fig:fexafs}}
\end{figure*}

Of relevance to the cross sections presented in this paper, 18
measurements (9 mounted on indium substrate and 9 on carbon)  of each
sample were taken. However, the carbon substrate measurements were
consistently nosier so we exclude those from the final co-added cross
sections.  For each individual  indium substrate cross section measure
(see Fig.~\ref{fig:normcompare}-top), the data were normalized  using
the aforementioned Eqs.~\ref{eq:bkg}~and~\ref{eq:norm} technique based
on the subsequent steps.   
To compare with our reference spectrum, each measurement is
first shifted by 4.1 eV to correct for the calibration offset of the
monochromator. The resultant energy span of our measurements 
is then 684.1~eV to 754.1~eV which corresponds to $N=700$ total data
bins of 0.1eV widths.   
As described, Eqs.~\ref{eq:bkg}~and~\ref{eq:norm} were then used to determine the
best normalized description for the combined data set unique to each
sample.  Since all data measured spanned the range noted above, the
number of pre-edge and post-edge data bins correspond respectively to
$n_1=179$ (684.1~eV$-$702~eV) and $n_2=141$ (740~eV$-$754.1~eV) for all
samples. The edge energy $E_{\rm edge}$=707~eV corresponding to the
Fe~L~{\sc iii}, and a polynomial of order $m=2$ was deemed sufficient, i.e.
the goodness of fit showed negligible improvement for higher polynomial orders.
Post-normalization,  cross sections of the highest quality were
co-added to create the final cross sections of Fig.~\ref{fig:fexafs}.
For the final cross sections, no fewer than 9 (indium substrate) data
sets were co-added per compound type.    Fig.~\ref{fig:fexafs} shows
our measured cross sections post-normalization, compared against
metallic \citep{kortrightkim00-fel}, and what we assume for bound-free
conintuum absorption by iron (\citealt{CL-abstables,henke}).  For future fitting efforts
to astrophysical data, we use the structure-less Fe~L tabulated values
of \cite{CL-abstables} to extrapolate the normalized XAFS data beyond
the energy span of our measurements to encompass an energy range that
extends from 0.4~keV$-10$~keV.

We plan to avail the community of our cross section
measurements\footnote{http://www.cfa.harvard.edu/hea/eg/isd.html}
{for these and other compounds, as absolutely
calibrated standards for interstellar dust studies, in the near future.  As stated,  the data
shown here is intended merely as a vehicle to facilitate a 
presentation of our new techniques.  }


\section{An X-ray method to determine the quantity and composition of
dust in interstellar space} \label{sec:gasfromdust} 
Our ability to accurately measure the quantity of dust and elemental abundances
in our Galaxy and beyond has far-reaching applications and
consequences for a diverse range of astrophysical topics.   Thus far,
wavelength-dependent  (IR to UV) studies of starlight attenuation as
facilitated by  e.g. E(B-V) measurements, and other extinction
studies,  have been the primary technique  by which we come to measure
the amount of dust that is bound up in interstellar grains.  Elemental
depletion can be determined by UV absorption studies comparing the
amount expected in gas phase absorption from what is observed, and IR
spectral studies can directly measure certain ISM dust.  Here, we 
present an {\it X-ray technique for directly determining the (element-specific)
quantity and composition of interstellar gas (\S\ref{sec:ismgas}) and dust from  
within a single observation.} What knowledge we gain from this
technique, when combined with the wealth of knowledge from non-X-ray
studies will significantly increase our understanding of ISM dust and
its effects on astrophysical environments.  

\subsection{Gas Phase ISM Absorption}  \label{sec:ismgas}
Like  condensed material, atomic transitions to higher quantum
levels within an isolated atom will also give rise to multiplet
resonant absorption.  To give the example for Fe, these lines would consist of all
possible discrete transitions to all possible configurations 
of the 3d-shell, since the electronic configuration of $Z$=26~$Fe^{+0}$
(or \ion{Fe}{1} in the astronomy notation) is $1s^2 2s^2 2p^6 3s^2 3p^6 4s^2 3d^6$.  
Therefore, while the bound-free non-resonant  transitions can be modelled by a simple step function
as e.g. that of \cite{CL-abstables} for Fe~L, 
additional discrete resonant features will have to be included
to account for the bound-bound resonant component of absorption. 

Fig.~\ref{fig:fegas} shows our calculations for the resonant 
transition for various Fe ions from $Fe^{+0}-Fe^{+4}$, as 
evaluated  based on the
Gu et al.~(2006\nocite{gu06fe}; see also \citealt{gu05fe})
predictions for oscillator strengths, radiative decay rates,
and autoionization rates for these lines.   
We note that because of the electronic structure of Fe,
which favors the removal of the s-electrons prior to d-electrons,
the cross sections for $Fe^{+0}\sim Fe^{+1}\sim Fe^{+2}$,
as evident in the figure.
The relevant cross sections as
derived for these atomic transitions
are then convolved with the appropriate instrument resolution
($\Delta E\sim0.9$~eV at 700~eV Fe~L for the Chandra Medium Energy Grating) to
approximate the ion-specific cross section for the bound-bound transition, $\sigma_{bb}$.
Finally, the cross section for describing the {\it total} gas-phase absorption for the species
of interest, $\sigma_{Zgas}$, will be described by the combination of cross-sections from
the bound-bound ($\mu_{bb-atom}$) and bound-free ($\mu_{bf}$) components.
More details on this follow in \S\ref{sec:g2d}.

For complex astrophysical environments, additional considerations as relates
to the ionization state of the plasma being probed need to be weighed.
As such, modelling efforts need to consider the  ionization state
of  the environment  expected for the absorption.   
For the three, cold~($T < 200$K), warm~($T\sim 8000$K), and 
hot~($T\sim 5\times10^5$K), phases of the ISM, the iron ions that
contribute most strongly to the L~edge region are 
$Fe^{+0}$ followed by $Fe^{+1}$. At 
T=8000~K,  the peak ion fractions are $Fe^{+0}\sim0.67$,
$Fe^{+1}\sim0.33$, $Fe^{+2}\sim2.8\times10^{-3}$
and $Fe^{+3}\sim1.9\times10^{-6}$, assuming the ISM ionizing
spectrum defined in \cite{sternberg02:ism}.
At the much colder temperatures of the cold ISM, only $Fe^{+0}$
contributes, and at the much hotter temperatures of the hottest phase
of the ISM, it is $Fe^{+25}$ (at 6.7~keV) which contributes the bulk
ion fraction.

\subsection{A prescription for determining species-specific gas-to-dust ratios} \label{sec:g2d}
\begin{figure*}
\includegraphics[width=0.80\textwidth, angle=-90]{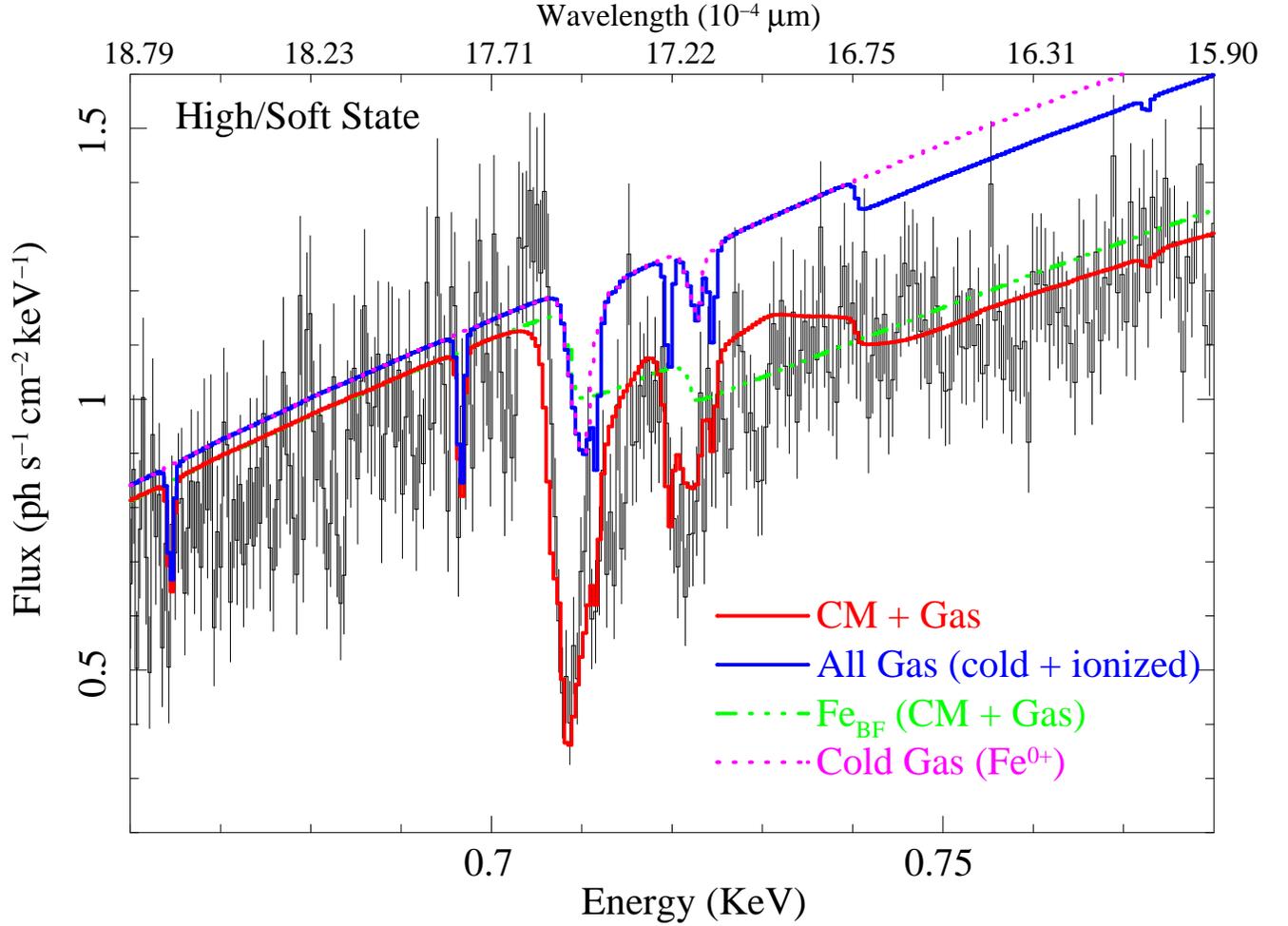}
\caption{The 700~eV Fe~L photoelectric edge region as observed in the 
X-ray binary Cygnus X-1 reveals how we can decompose the absorption components 
to determine the quantity and composition of dust.  The best fit
(red) shows the combined absorption from all gas (blue) which includes 
line-of-sight cold gas (magenta) and XAFS signifying condensed matter~(CM).  Here, 
the ionized gas component contribution comes largely from plasma that
is optically thick to He-like \ovii, and therefore the $\rm 1s^2-1s$$n$p
(for $n > 3$) \ovii resonance lines (blue) contribute to the absorption region 
co-located with the Fe~L{\sc iii} and L~{\sc ii} edges.  In green is the 
absorption component representing the bound-free transition from the Fe~ion
{\it and} solid/molecule. 
\label{fig:cygx1}}
\end{figure*}

As stated in \S\ref{sec:intro}, what  cross sections we measure in the
laboratory of astrophysically-likely dust candidates
(\S\ref{sec:lab}) will be applied  to X-ray spectra showing structure
near photoelectric edges to determine  dust composition,  {\it and}
quantity.  We propose the following prescription to do so.

\begin{equation}
F_{\rm final} = F_0\,T = F_0\, \rm exp\,[-\tau_{\rm LOS(Z')} - \tau_{ionized-gas} - \tau_{\tiny (Zgas)} - \tau_{\small (Zdust)}]
\end{equation} 
whereby the absorption contributions from gas (cold and ionized), and
dust, can be {\it separately} accounted for by each of the exponential
terms.   $F_{\rm final}$ and  $F_0$ are respectively the detected and 
incident flux, and $T$,  represented by the exponential terms is
the transmission.
$\tau_{ionized-gas}$ is determined through photoionization modelling
to account for any additional lines arising from the plasma of the
source, and should be treated on a source by source basis. However, 
in the Fe~L{\sc iii} and L~{\sc ii} edge regions,
we note additional line  contributions would have to  come from high transition 
(low oscillator strength) lines of He-like \ion{O}{7} ($O^{+6}$) $1s^2-1snp$
for $n \ge 4$, and \ion{Fe}{1}-\ion{Fe}{11} (i.e. $Fe^{+0} - Fe^{+10})$.
Any significant contribution from oxygen would have to come from
very ionized optically thick plasma; for the iron, moderate temperature plasma
would have to be present in large quantities to have any appreciable effect in
this spectral region.  Therefore, for many astrophysical situations, neither ionized oxygen nor iron from the 
source plasma, is expected to 
contribute strongly (if at all) to the Fe~L~{\sc iii} and Fe~L~{\sc ii} spectral
region. Nevertheless, as stated, these can be modelled out through photoionization studies,
the discussion of which is beyond the scope of this paper.  
For the total line-of-sight (LOS) (gas+dust) ISM absorption from all heavy elements,
{\it excluding} the species of interest, 
\begin{equation} \label{eq:Nx}
\tau_{\rm LOS(Z')} = \sum_{Z'} \sigma^{'}_{Z'} \, N_{\rm H} \, A_{\rm solar} 
\end{equation} 
where $\sigma_{Z'}$,  $A_{solar}$, and $N_{\rm H}$ are
respectively defined to be  the cross section, abundance, and
line-of-sight equivalent Hydrogen column summed over all heavy elements 
present in the broad band X-ray spectrum, {\it with the exception} of the element we are trying
to measure. One reason for removing the species of interest from the
broad-band fitting is because we are interested in decomposing  the
gas and dust contribution specific to that species.  For this, we
include two additional exponential terms to describe the 
absorption by the gas and dust for the species of interest
whereby:
\begin{equation}  \label{eq:nzgas}
\tau_{\rm Zgas} = {(\sigma_{bf}+\sigma_{bb-gas})  N_{\rm H} xA_Z} = {\sigma_{Zgas} N_{\rm H} xA_Z}  
\end{equation}
and 
\begin{equation}  \label{eq:ndust}
\tau_{\rm Zdust} =  (\sigma_{bf}+\sigma_{bb-dust}) N_{\rm H} (1-x) A_Z = \sigma_{Zdust} N_{\rm H} (1-x) A_Z
\end{equation} 
In these equations,  $\sigma_{Zgas}$ and $\sigma_{Zdust}$ refer to the
cross section of the gas and dust respectively for the species of
interest. The determination of $\sigma_{Zgas}$ was discussed in \S\ref{sec:ismgas}.  
$\sigma_{Zdust}$, and our efforts to measure this for
astrophysically viable dust candidates has been discussed at length 
in \S\ref{sec:lab}.  
 Given enough S/N, the species-specific abundance $A_Z$ can be fitted for,
with the condition that the total abundance should be a combination of
both gas and dust, hence the additional multiplicative factor to $A_Z$
of $x$ and $(1-x)$.  
Fig.~\ref{fig:cygx1} shows these components as separated
out in a fit to the X-ray binary Cygnus~X1.  $-$~Details about the
dust in Cygnus~X1 will be separately discussed in the paper
\citealt{lee:cygx1dust}.


\section{Discussion} \label{sec:discussion} 
The Chandra and XMM-Newton archive are rich with high resolution
spectra  of X-ray bright binaries and active galactic nuclei (AGN).  
A cursory look at these
spectra reveal XAFS near many photoelectric edges, and hence dust
and/or molecules.  Therefore, there is already a wealth of data for
studying astrophysical  dust properties in the ISM and/or the hotter
``near binary'' environments using the X-ray technique described in
this paper.  In combination with X-ray scattering studies which will
provide important spatial information on distribution (see
e.g. \citealt{xiang07:halo} paper which make a first  successful
attempt at considering non-uniform dust distributions along the
line-of-sight toward the X-ray dipper \1624, and references therein),
existing grating observations provide us with great potential to map
out dust distribution along different lines-of-sight, while
determining the species-specific quantity and composition along these
same path-lengths to  build upon studies in other wavebands.   Given
spectra with enough S/N (as true of many of the existing data),   the
spectral resolution afforded by the Chandra gratings will allow us to
(1) definitively separate dust from gas phase absorption, and (2) give
us good expectations for being able to {\it directly} determine the chemical
composition of dust grains containing oxygen, magnesium, silicon,
sulfur, and iron.  As such, we expect to be able to make significant
progress over the next few years in our understanding of the
sub-micron size molecules which make up a large fraction of the ISM
grains, and/or near-binary/near-AGN environments.  For the latter,
this would enable a better understanding of the evolutionary histories
of these black hole and neutron star systems.

For many astrophysical applications,  a 
better understanding of dust properties and its
distribution in interstellar space, carries many advantages.  As an example,
consider  the supermassive black hole Sagittarius~A$^\ast$  in our own Galactic center.  Current
estimates of the absorption column to Sagittarius A$^\ast$ between
X-ray and NIR studies differ by a factor of $\sim$~2,  which may be attributed to
unknowns relating to the metallicity, gas-to-dust ratio, or the  grain
size distribution along the line of sight.  These uncertainties
affect our ability to accurately use the Fe~K$\alpha$ emission line
as a diagnostic of the Sagittarius~A$^\ast$ hot accretion flow (see
e.g. \citealt{xu06}),  and determination of the X-ray spectral slope
for observed flares.  Metallicity uncertainties can be greatly
improved as we understand the nature of the absorption towards the
Galactic center better.  This  can be accommodated by the type of
study described in  this paper, in large part because the spatial
distribution of sources with spectra of sufficient S/N and spectral
resolution necessary for the proposed  X-ray study of dust, encompass
much of the region along and around the Galactic plane, and towards
the Galactic center.     Ideally,   we would additionally conduct case
studies of the fainter binaries in the central parsecs, or at least as
many as possible, as close to the Galactic center as possible  (see
e.g. \citealt{muno08} latest X-ray binary catalog based on \chandra
ACIS-I imaging data).    Given the relative X-ray faintness of many of
these objects however, such a comprehensive spectral study in the
nearest regions of the  Galactic Center will have to await missions
with higher throughput, and at least equal spectral resolution.

As stated, studies of many astrophysical systems can be greatly
enhanced by a better understanding of the line-of-sight absorbing
material.  With the \chandra gratings, we have entered upon an era
where condensed matter theory can be applied to astrophysics studies
of dust.  Invariably, interstellar environments present us with
additional complications not of consideration in a {\it controlled}
laboratory experiment, so we conclude with a thought problem of
relevance to dust studies of interstellar environments.

\subsection{A Gedanken problem: can we differentiate dust content in different locations?}
\label{sec:gedanken} 
The path-length along the line-of-sight to our illuminating sources,
be they X-ray binaries in our own Galaxy or black hole systems in
other galaxies, is complex.  Consider, for example, the simplified
scenario of only two components distributed in some manner along the
path.  This problem then has two parts, the identification of the
chemical species of those two components and the determination of
their distribution along the path-length.  Assuming we can identify
the two species, can we distinguish a heterogeneous arrangement~(dust
of composition~a in location~A plus dust of composition~b found in
location~B), from a homogeneous arrangement~(dust of mixed composition
AB distributed along the path) ?

The interpretation of the X-ray spectrum passing along this
line-of-sight bears a strong similarity to the well-established
technique of XAS as performed at
synchrotron facilities.  The use of XAS to identify chemical species
in a chemically heterogeneous sample is common practice in terrestrial
laboratories  (\citealt{lengke06} is but one example among
thousands in the XAS literature.)  The common strategy is to interpret
the spectrum from the mixture of species as a linear combination of
the spectra from its component species, i.e.  
$\displaystyle\sum_{i=0}^{n} c_i \alpha_{i}$, where $\alpha_{i}$ represent the 
different species and $c_i$ are their respective fractional contributions,
as per \cite{lengke06}. The best determined relative percentages for the linear combination are derived by fits based on minimal $\chi^2 $,  which we recast for our purposes to be 
\begin{equation} \label{eq:lincombochi}
\sum_{j=1}^m\left[{\sum_{i=1}^n c_i\, \alpha_{i,j} - \beta_j \over \sqrt{\sum_{i=1}^n (c_i\, \Delta \alpha_{i,j})^2 + (\Delta \beta_j)^2}} \right]^2
\end{equation}
where $\beta_j$ and $\alpha_{i, j}$ are the normalized cross sections in the $j$th bin of respectively the mixture and pure forms of the individual compounds, $\Delta\beta_j$ and $\Delta\alpha_{i, j}$ their respective uncertainties,  and $n$ the associated number of compounds and $j$th bins to sum over;
$c_i$ has the condition that $\displaystyle\sum^n_{i=1} c_i$ = 1. 
With careful sample preparation, the fractional content of the
component species in the spectrum is directly indicative of their
fractional content in the physical sample.

To demonstrate this using iron-bearing species of relevance to the
ISM, we performed the following experiment in the controlled
laboratory of the synchrotron facility.  We prepared two samples
containing the common terrestrial iron compounds hematite
($\alpha$-Fe$_2$O$_3$), ferrosilicate (Fe$_2$SiO$_4$), and lepidocrocite
($\gamma$-FeOOH).  The first sample contained equal parts by weight of
hematite and ferrosilicate and the second contained equal parts by
weight of all three materials.  Samples for measurement by electron
yield at beamline 6.3.1 were prepared as described above and the XAS
spectra were measured on both mixtures.  The hematite and lepidocrocite
were very fine-grained, commercially produced powders.  The
ferrosilicate was commercially produced, but was ground by agate
mortar and pestle from a chunk several millimeters across.
Consequently, the ferrosilicate component of each sample was much
larger-grained than the other two components.

Because equal parts by weight were mixed in our two samples, we might
expect the measured spectra to be interpreted by equally weighted
fractions of the spectra from the three pure materials.  In fact the
nature of the interaction between the X-rays and the morphology of the
sample must be considered.  Because the ferrosilicate component of
each sample was much larger grained than the other components, its
surface-area-to-volume ratio was considerably smaller.  Because the
electron yield measurement is surface sensitive, a much smaller
fraction of the iron atoms in the ferrosilicate contributed to the
actual measurement than of the other two components.

\begin{figure}
\includegraphics[width=0.85\textwidth, angle=0]{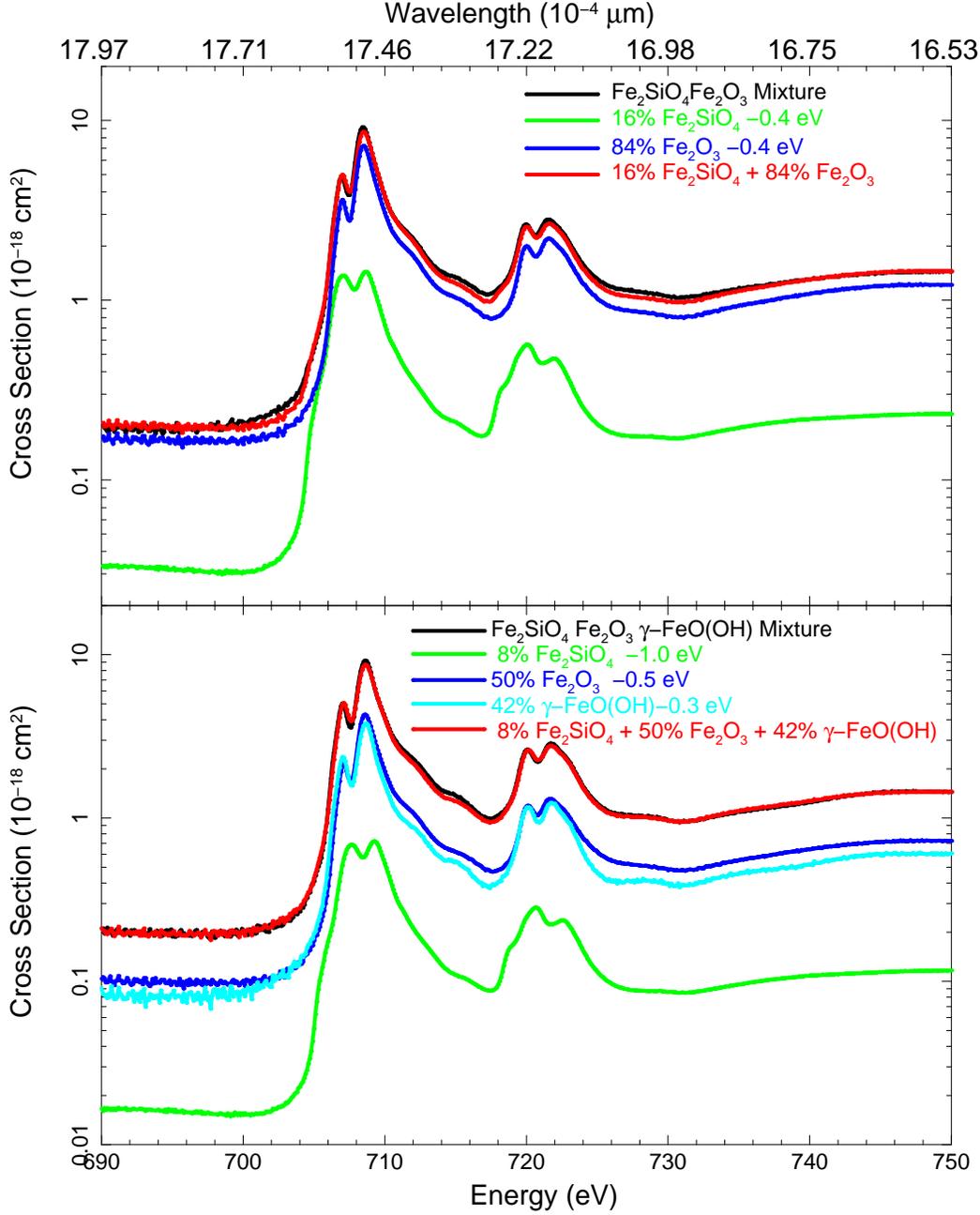}
\caption{A comparison of the linear combination (red) of the pure compounds against the mixture (black) for the problem considered in \S\ref{sec:gedanken}.  In blue and green are the cross sections of the individual compounds  reduced according to their fractional contributions in the linear combination;  small energy shifts are allowed.   The best determined relative percentages for the linear combination are derived by fits based on minimal $\chi^2 $ according to Eq.~\ref{eq:lincombochi}. This color coding applies to both the linear combination of the binary (top) and trinary (bottom) mixtures.
\label{fig:lincombo}
}
\end{figure}

The data on the two mixtures along with the results of the linear
combination analysis are shown in Fig.~\ref{fig:lincombo}.  The binary
mixture proved to be $16\pm2$\% ferrosilicate and $84\pm2$\% hematite.
As expected from particle size, the ferrosilicate is under-represented
in the measured spectrum.  The trinary mixture was $8\pm2\%$
ferrosilicate, $51\pm7$\% hematite, and $41\pm7$\% lepidocrocite.  The two
components of similar particle size are evenly represented, while the
ferrosilicate is under-represented.

This careful consideration of sample morphology in the synchrotron
measurement is directly relevant to the interpretation of the
astrophysical X-ray spectra.  The astrophysical measurement is akin to
a transmission measurement at the synchrotron, while the examples
shown in Fig.~\ref{fig:lincombo} were measured in electron
yield. Still, the correct interpretation of the XAS spectra requires
consideration of the nature of the interaction of the X-rays with the
material through which they pass.  A ray interacting with a large
particle in the ISM will be lost to the satellite measurement by
virtue of being completely absorbed by the particle.  Consequently,
the satellite measurement will be dominated by the small particles in
the ISM as those are the only particles that allow passage of photons
for eventual measurement at the satellite.

Finally, we must address the topic of distinguishing the distribution
of the chemical species along the path-length.  By itself, a
transmission XAS measurement is incapable of distinguishing between a
layered and a homogenized sample.  In transmission, an XAS measurement
on two powders will be the same whether you prepare the powders
separately and stack them in the beam path or mix the powders
thoroughly before making the measurement.  Similarly, the X-ray
satellite measurement by itself cannot address the distribution of the
species it is able to identify, {\it if} co-located at similar redshifts.  
However, { since an XAFS contribution to a photoelectric edge is 
scaled according to the optical depth of that particular species, we can
reasonably conclude that the compound that best fits the astro- spectra
is the compound where we are seeing the bulk contributions from, even if there
might be lesser (e.g. 10\%) contributions from other compounds along the line-of-sight.}
Therefore, in combination with other
astrophysical studies  using photon wavelengths that interact
differently with the ISM (e.g. IR), the X-ray measurement can serve an
important role in the full determination of the composition of the
ISM.

\section{XAFS Science: Present and Future} \label{sec:future} 
The X-ray energy band has been slow to be exploited for dust studies
due largely to instrumental requirements for both good spectral
resolution ($R \approxgt 1000$), and throughput.   Yet, {\it condensed
matter astrophysics} (i.e. the merging of high energy condensed matter
and astrophysics techniques for X-ray studies of dust), as a new
sub-field can be realized in the present era of \chandra and \XMM.  As
such, X-rays should be considered a powerful and viable new resource
for delving into a relatively unexplored regime for determining dust
properties: composition, quantity, and distribution.  Present day
studies with extant satellites will set the foundation for future
studies with larger, more powerful missions such as the joint ESA-JAXA-NASA
 {\it International X-ray Observatory} (\footnote{http://ixo.gsfc.nasa.gov/}{\it
IXO}). Proposed  spectral
instruments such as the Critical-Angle Transmission grating
(\footnote{http://space.mit.edu/home/dph/ixo/}$^,$\footnote{http://space.mit.edu/home/dph/ixo/comparison.html}CAT; \citealt{CAT07})
and Off-Plane Reflection Gratings \citep{ixooffplane} are designed to provide {\it IXO's}
baseline resolving power of 3000 under 1 keV, and configurations are contemplated which will boost
this to 5000 or more.  Similar baseline resolving power are also expected from the 
calorimeters for  the hard ($> 6$~keV) spectral region. 
{\it IXO's} target spectral resolution $R=3000$ in the 0.2-10~keV bandpass,
in  combination with planned higher throughput (10$\times$ XMM, and 60$\times$ Chandra) 
will allow us in the future to move beyond the realm discussed in this
paper to being able to use  XAFS to recreate the crystalline structure
of interstellar grains and determine precise oxidation states. 
In the interest of providing information for the planning of future missions,
we provide spectral resolution goals that are mapped to XAFS science 
hurdles, for gratings and calorimeters (Table~\ref{tab:specres}).

\section*{acknowledgements}
We acknowledge Eric Gullickson, Pannu Nachimuthu, and Elke Arenholz  for beamline support.
We thank Claude Canizares and Alex Dalgarno for advice and conversations, 
and  Fred Baganoff for discussions relating to Sgr~A$^\ast$.
The Advanced Light Source and JKB is supported by the Director, Office of Science, 
Office of Basic Energy Sciences, of the U.S. Department of Energy under Contract 
No. DE-AC02-05CH11231.   JCL is grateful to Chandra grant
SAO AR8-9007 and  the Harvard Faculty of Arts and Sciences for financial support.

\begin{deluxetable}{lccccc}
\tablecaption{Photoelectric edge energies of measured compounds.}
\tablecolumns{6}
\tablehead{\colhead{Compounds$^a$} & \colhead{Charge$^b$} & \colhead{E$\rm_{LIII}^c$} & \colhead{$\rm \Delta E_{LIII}^d$} & \colhead{\rm $\rm E_{LII}^e$}  & \colhead{$\rm \Delta E_{LII}^f$}\\
\colhead{} & \colhead{state} & \colhead{eV($\rm \AA$)} & \colhead{eV($\rm \AA$)} & \colhead{eV($\rm \AA$)} & \colhead{eV($\rm \AA$)} }
\startdata
FeO(OH)-lepidocrocite  & Fe$^{3+}$  & 705.7(17.57)  &  -2.7 (+0.07)  & 718.9(17.25)  & -2.3 (+0.06)   \\
Fe$_2$O$_3$-hematite  & Fe$^{3+}$  & 705.7(17.57)  &  -2.7 (+0.07)  & 718.9(17.25)  & -2.3 (+0.06)  \\
Fe$_2$SiO$_4$-fayalite & Fe$^{2+}$  & 704.3(17.60)  &  -4.1 (+0.10)  & 717.4(17.28)  & -3.8 (+0.09) \\
FeSO$_4$-iron sulfate  & Fe$^{2+}$  & 704.3(17.60)  &  -4.1 (+0.10)  & 717.8(17.27)  & -3.4 (+0.08) \\
\enddata
\scriptsize
\tablenotetext{a}{~Laboratory measurements of Cross sections for astrophysically-likely species presented in this paper.}
\tablenotetext{b}{~The charge state reflects the number of electrons removed from iron when the compound is formed.  To take Fe$_2$SiO$_4$ (fayalite) as an example: silicon (atomic orbital: $1s^22s^22p^63s^23p^2$), will lose 4 electrons in its outmost shell so that oxygen ($1s^22s^22p^4$) can fill its shell.  Since there are 4 oxygen atoms however, two additional electrons will be taken from Fe such that the charge state
of the compound is +2.}
\tablenotetext{c}{~The Fe~$\rm L_{III}$ photoelectric edge energy relevant to the different compounds as measured at where the \cite{CL-abstables} cross sections, which we assume for bound-free gas phase iron absorption, peaks at 708.4 eV.  See Fig.~\ref{fig:fexafs} inset for illustration.}
\tablenotetext{d}{~The energy difference between compound and gas as measured at the peak of the \cite{CL-abstables}  } 
\tablenotetext{e}{~Same as column c but for the Fe~$\rm L_{II}$ photoelectric edge at 721.2 eV assumed for bound-free gas phase absorption.}
\tablenotetext{f}{~Same as column d but for the Fe~$\rm L_{II}$ photoelectric edge at 721.2 eV.}
\label{tab:feledges}
\end{deluxetable}

\begin{table}
   \tablecaption{X-ray Instrument Capabilities and  Goals for ISM Grain Physics Studies}
 \begin{threeparttable}
   \caption{X-ray Instrument Capabilities and Goals for ISM Grain Physics Studies}
  \label{tab:specres}
   \begin{tabular}{|l|c|c|c|c|c|}
     \cline{1-6}
           & \multicolumn{2}{|c|}{Gratings}        & \multicolumn{2}{|c|}{Calorimeter}   & \\
           & \multicolumn{2}{|c|}{0.25-6~keV FWHM} & \multicolumn{2}{|c|}{6-10~keV FWHM} & \\
     \cline{2-5}

           &            & Chandra
           &  Suzaku$^\dagger$                   &    SXS prototype$^\ddagger$             & IDEAL$^\S$                  \\ 
     Spectral Resolution$^\star$      & $ R < 500 $ & $R=1000$
           & $R=923$ & $R \sim1500$ & $R$ \\ 
     \hline
     Distinguish gas from dust & difficult\tnote{1} & yes\tnote{2} & yes         & yes            &      \\ 
     \hline
     Differentiate dust        & not possible       & yes\tnote{3} & ok\tnote{3} & yes\tnote{3}   & 3000\tnote{4} \\ 
     \hline
     Discern oxides            & not possible       & no           & no          & maybe\tnote{5} & 5000 \\ 
     \hline
    \end{tabular}
    \begin{tablenotes}
	\item[$\star$] Spectral resolution $R =  \rm{E/\Delta E} = \rm{\lambda/\Delta \lambda}$ referenced to 1 keV for the gratings at 6 keV for the calorimeters.
	\item[$\dagger$] The spectral resolution of the calorimeter on board Suzaku which has since failed in-flight.
	\item[$\ddagger$] Reasonable expectation for spectral resolution to be achieved with SXS prototype detectors (HgCdTe absorbers and implanted silicon thermometers) planned for the Astro-H mission.  4~eV resolution ($R=1500$) at 6~keV has been achived in laboratory measurements. -- Astro-H Team (private communication). The IXO calorimeter baseline performance is 2.5 eV. 
	\item[\S] Applies to the 0.25-10~keV range for both grating and calorimeter instrument considerations.
      \item[1]{Insufficient spectral resolution limits our ability to  discern the component of the 
photoelectric edge due to dust versus gas.}
      \item[2]{As stated in \S\ref{sec:intro}, the soft X-ray band is also complicated by ionized absorption lines imprinted by the  hot plasma environment of the illuminating source (black hole or neutron star, e.g.)  that need to be  modeled out in order to isolate the XAFS features.}
      \item[3]{Possible with adequate statistics obtained either from long observation time or large measurement area}
      \item[4]{Different forms of iron (silicates, oxides, metal) and be distinguished, but not oxides cannot be distinguished from other oxides.}
      \item[5]{Challenging measurement requiring measurement statistics approaching the level of synchrotron experiment}
    \end{tablenotes}
  \end{threeparttable}
\end{table}

\clearpage


\clearpage

\clearpage

\clearpage

\begin{figure*}
\includegraphics[height=0.70\textheight, angle=0]{Fe_I-V.eps}
\caption{The bound-bound resonant transition for various Fe ions
from $Fe^{+0}-Fe^{+4}$,  as calculated based on the
\citep{gu06fe,gu05fe} predictions for oscillator strengths, radiative decay rates,
and autoionization rates for these lines.   Based on the 
ISM ionization spectrum of \cite{sternberg02:ism}, the most prominent 
contribution from Fe ions in the L-edge region would come from
neutral $Fe^{+0}$ (i.e. \ion{Fe}{1}), or singly ionized $Fe^{+1}$ (\ion{Fe}{2}).
These lines are convolved with the Chandra HETGS spectral resolution
($R\sim0.9$~eV at FeL;  blue), and the resolution of ALS beamline~6.3.1 used
for the XAFS measurements presented in this paper (red).  At present,
$R\sim3000$ is also the baseline spectral resolution for the IXO spectrometers.
\label{fig:fegas}}
\end{figure*}

\clearpage

\end{document}